# Why Neurons Have Thousands of Synapses, A Theory of Sequence Memory in Neocortex

Jeff Hawkins*, Subutai Ahmad

Numenta, Inc, Redwood City, California, United States of America

*Corresponding author
Emails: jhawkins@numenta.com, sahmad@numenta.com

A version of this manuscript has been submitted for publication as of October 30, 2015. *Note that figures and tables are at the end of this PDF.*

Please contact the authors for updated citation information.

# Abstract

Neocortical neurons have thousands of excitatory synapses. It is a mystery how neurons integrate the input from so many synapses and what kind of large-scale network behavior this enables. It has been previously proposed that non-linear properties of dendrites enable neurons to recognize multiple patterns. In this paper we extend this idea by showing that a neuron with several thousand synapses arranged along active dendrites can learn to accurately and robustly recognize hundreds of unique patterns of cellular activity, even in the presence of large amounts of noise and pattern variation. We then propose a neuron model where some of the patterns recognized by a neuron lead to action potentials and define the classic receptive field of the neuron, whereas the majority of the patterns recognized by a neuron act as predictions by slightly depolarizing the neuron without immediately generating an action potential. We then present a network model based on neurons with these properties and show that the network learns a robust model of time-based sequences. Given the similarity of excitatory neurons throughout the neocortex and the importance of sequence memory in inference and behavior, we propose that this form of sequence memory is a universal property of neocortical tissue. We further propose that cellular layers in the neocortex implement variations of the same sequence memory algorithm to achieve different aspects of inference and behavior. The neuron and network models we introduce are robust over a wide range of parameters as long as the network uses a sparse distributed code of cellular activations. The sequence capacity of the network scales linearly with the number of synapses on each neuron. Thus neurons need thousands of synapses to learn the many temporal patterns in sensory stimuli and motor sequences.

## 1. Introduction

Excitatory neurons in the neocortex have thousands of excitatory synapses. The proximal synapses, those closest to the cell body, have a relatively large effect on the likelihood of a cell generating an action potential. However, a majority of the synapses are distal, or far from the cell body. The activation of a single distal synapse has little effect at the soma, and for many years it was hard to imagine how the thousands of distal synapses could play an important role in determining a cell's responses (Major et al., 2013). We now know that dendrite branches are active processing elements. The activation of several distal synapses within close spatial and temporal proximity can lead to a local dendritic NMDA spike and consequently a significant and sustained depolarization of the soma (Antic et al., 2010; Major et al., 2013). This has led some researchers to suggest that dendritic branches act as independent pattern recognizers (Poirazi et al., 2003; Polsky et al., 2004). Yet, despite the many advances in understanding the active properties of dendrites, it remains a mystery why neurons have so many synapses and what their precise role is in memory and cortical processing.

Lacking a theory of why neurons have active dendrites, almost all artificial neural networks, such as those used in deep learning (LeCun et al., 2015) and spiking neural networks (Maass, 1997), use artificial neurons without active dendrites and with unrealistically few synapses, strongly suggesting they are missing key functional aspects of real neural tissue. If we want to understand how the neocortex works and build systems that work on the same principles as the neocortex, we need an understanding of how biological neurons use their thousands of synapses and active dendrites. Of course, neurons cannot be understood in isolation. We also need a complementary theory of how networks of neurons, each with thousands of synapses, work together towards a common purpose.

In this paper we introduce such a theory. First, we show how a typical pyramidal neuron with active dendrites and thousands of synapses can recognize hundreds of unique patterns of cellular activity. We show that a neuron can recognize hundreds of patterns even in the presence of large amounts of noise and variability as long as overall neural activity is sparse. Next we introduce a neuron model where the inputs to different parts of the dendritic tree serve different purposes. In this model the patterns recognized by a neuron's distal synapses are used for prediction. Each neuron learns to recognize hundreds of patterns that often precede the cell becoming active. The recognition of any one of these learned patterns acts as a prediction by depolarizing the cell without directly causing an action potential. Finally, we show how a network of neurons with this property will learn and recall sequences of patterns. The network model relies on depolarized neurons firing quickly and inhibiting other nearby neurons, thus biasing the network's activation towards its predictions. Through simulation we illustrate that the sequence memory network exhibits numerous desirable properties such as on-line learning, multiple simultaneous predictions, and robustness.

Given the similarity of neurons throughout the neocortex and the importance of sequence memory for inference and behavior, we propose that sequence memory is a property of neural tissue throughout the neocortex and thus represents a new and important unifying principle for understanding how the neocortex works.

## 2. Results

### 2.1. Neurons Recognize Multiple Patterns

It is common to think of a neuron as recognizing a single pattern of activity on its synapses. This notion, sometimes called a "point neuron", forms the basis of almost all artificial neural networks (Fig. 1A).

[Figure 1 about here – see end of manuscript]

Active dendrites suggest a different view of the neuron, where neurons recognize many unique patterns (Larkum and Nevian, 2008; Poirazi et al., 2003; Polsky et al., 2004). Experimental results show that the coincident activation of eight to twenty synapses in close spatial proximity on a dendrite will combine in a non-linear fashion and cause an NMDA dendritic spike (Larkum et al., 1999; Major et al., 2013; Schiller and Schiller, 2001; Schiller et al., 2000). Thus, a small set of neighboring synapses acts as a pattern detector. It follows that the thousands of synapses on a cell's dendrites act as a set of independent pattern detectors. The detection of any of these patterns causes an NMDA spike and subsequent depolarization at the soma.

It might seem that eight to twenty synapses could not reliably recognize a pattern of activity in a large population of cells. However, robust recognition is possible if the patterns to be recognized are sparse; i.e. few neurons are active relative to the population (Olshausen and Field, 2004). For example, consider a population of 200K cells where 1% (2,000) of the cells are active at any point in time. We want a neuron to detect when a particular pattern occurs in the 200K cells. If a section of the neuron's dendrite forms new synapses to just 10 of the 2,000 active cells, and the threshold for generating an NMDA spike is 10, then the dendrite will detect the target pattern when all 10 synapses receive activation at the same time. Note that the dendrite could falsely detect many other patterns that share the same 10 active cells. However, if the patterns are sparse, the chance that the 10 synapses would become active for a different random pattern is small. In this example it is only $9.8 \times 10^{-21}$.

The probability of a false match can be calculated precisely as follows. Let $n$ represent the size of the cell population and $a$ the number of active cells in that population at a given point in time, for sparse patterns $a \ll n$. Let $s$ be the number of synapses on a dendritic segment and $\theta$ be the NMDA spike threshold. We say the segment recognizes a pattern if at least $\theta$ synapses become active, i.e. at least $\theta$ of the $s$ synapses match the currently active cells.

Assuming a random distribution of patterns, the exact probability of a false match is given by:

$$\frac{\sum_{b=\theta}^{s} \binom{s}{b} \times \binom{n-s}{a-b}}{\binom{n}{a}} \qquad (1)$$

The denominator is simply the total number of possible patterns containing $a$ active cells in a population of $n$ total cells. The numerator counts the number of patterns that would connect to $\theta$ or more of the $s$ synapses on one dendritic segment. A more detailed description of this equation can be found in (Ahmad and Hawkins, 2015).

The equation shows that a non-linear dendritic segment can robustly classify a pattern by sub-sampling (forming synapses to only a small number of the cells in the pattern to be classified). Table A in S1 Text lists representative error probabilities calculated from Eq. (1).

By forming more synapses than necessary to generate an NMDA spike, recognition becomes robust to noise and variation. For example, if a dendrite has an NMDA spike threshold of 10, but forms 20 synapses to the pattern it wants to recognize, twice as many as needed, it allows the dendrite to recognize the target pattern even if 50% of the cells are changed or inactive. The extra synapses also increase the likelihood of a false positive error. Although the chance of error has increased, Eq. (1) shows that it is still tiny when the patterns are sparse. In the above example, doubling the number of synapses and hence introducing a 50% noise tolerance, increases the chance of error to only $1.6 \times 10^{-18}$.

Table 1B in S1 Text lists representative error rates when the number of synapses exceeds the threshold.

The synapses recognizing a given pattern have to be co-located on a dendritic segment. If they lie within 40μm of each other then as few as eight synapses are sufficient to create an NMDA spike (Major et al., 2008). If the synapses are spread out along the dendritic segment, then up to twenty synapses are needed (Major et al., 2013). A dendritic segment can contain several hundred synapses; therefore each segment can detect multiple patterns. If synapses that recognize different patterns are mixed together on the dendritic segment, it introduces an additional possibility of error by co-activating synapses from different patterns. The probability of this type of error depends on how many sets of synapses share the dendritic segment and the sparsity of the patterns to be recognized. For a wide range of values the chance for this type of error is still low (Table C in S1 Text). Thus the placement of synapses to recognize a particular pattern is somewhat precise (they must be on the same dendritic segment and ideally within 40μm of each other), but also somewhat imprecise (mixing with other synapses is unlikely to cause errors).

If we assume an average of 20 synapses are allocated to recognize each pattern, and that a neuron has 6,000 synapses, then a cell would have the ability to recognize approximately 300 different patterns. This is a rough approximation, but makes evident that a neuron with active dendrites can learn to reliably recognize hundreds of patterns within a large population of cells. The recognition of any one of these patterns will depolarize the cell. Since all excitatory neurons in the neocortex have thousands of synapses, and, as far as we know, they all have active dendrites, then each and every excitatory neocortical neuron recognizes hundreds of patterns of neural activity.

In the next section we propose that most of the patterns recognized by a neuron do not directly lead to an action potential, but instead play a role in how networks of neurons make predictions and learn sequences.

### 2.1.1. Three Sources of Synaptic Input to Cortical Neurons

Neurons receive excitatory input from different sources that are segregated on different parts of the dendritic tree. Fig. 1B shows a typical pyramidal cell, the most common excitatory neuron in the neocortex. We show the input to the cell divided into three zones. The proximal zone receives feedforward input. The basal zone receives contextual input, mostly from nearby cells in the same cortical region (Petreanu et al., 2009; Rah et al., 2013; Yoshimura et al., 2000). The apical zone receives feedback input (Spruston, 2008). (The second most common excitatory neuron in the neocortex is the spiny stellate cell; we suggest they be considered similar to pyramidal cells minus the apical dendrites.) We propose the three zones of synaptic integration on a neuron (proximal, basal, and apical) serve the following purposes.

*Proximal Synapses Define the Classic Receptive Field of a Cell*

The synapses on the proximal dendrites (typically several hundred) have a relatively large effect at the soma and therefore are best situated to define the basic receptive field response of the neuron (Spruston, 2008). If the coincident activation of a subset of the proximal synapses is sufficient to generate a somatic action potential and If the inputs to the proximal synapses are sparsely active, then the proximal synapses will recognize multiple unique feedforward patterns in the same manner as discussed earlier. Therefore, the feedforward receptive field of a cell can be thought of as a union of feedforward patterns.

*Basal Synapses Learn Transitions in Sequences*

We propose that basal dendrites of a neuron recognize patterns of cell activity that precede the neuron firing, in this way the basal dendrites learn and store transitions between activity patterns. When a pattern is recognized on a basal dendrite it generates an NMDA spike. The depolarization due to an NMDA spike attenuates in amplitude by the time it reaches the soma, therefore when a basal dendrite recognizes a pattern it will depolarize the soma but not enough to generate a somatic action potential (Antic et al., 2010; Major et al., 2013). We propose this sub-threshold depolarization is an important state of the cell. It represents a prediction that the cell will become active shortly and plays an important role in network behavior. A slightly depolarized cell fires earlier than it would otherwise if it subsequently receives sufficient feedforward input. By firing earlier it inhibits neighboring cells, creating highly sparse patterns of activity for correctly predicted inputs. We will explain this mechanism more fully in a later section.

*Apical Synapses Invoke a Top-down Expectation*

The apical dendrites of a neuron also generate NMDA spikes when they recognize a pattern (Cichon and Gan, 2015). An apical NMDA spike does not directly affect the soma. Instead it can lead to a Ca2+ spike in the apical dendrite (Golding et al., 1999; Larkum et al., 2009). A single apical Ca2+ spike will depolarize the soma, but typically not enough to generate a somatic action potential (Antic et al., 2010). The interaction between apical Ca2+ spikes, basal NMDA spikes, and somatic action potentials is an area of ongoing research (Larkum, 2013), but we can say that under many conditions a recognized pattern on an apical dendrite will depolarize the cell and therefore have a similar effect as a recognized pattern on a basal dendrite. We propose that the depolarization caused by the apical dendrites is used to establish a top-down expectation, which can be thought of as another form of prediction.

### 2.1.2. The HTM Model Neuron

Fig. 1C shows an abstract model of a pyramidal neuron we use in our software simulations. We model a cell's dendrites as a set of threshold coincidence detectors; each with its own synapses. If the number of active synapses on a dendrite/coincidence detector exceeds a threshold the cell detects a pattern. The coincidence detectors are in three groups corresponding to the proximal, basal, and apical dendrites of a pyramidal cell. We refer to this model neuron as an "HTM neuron" to distinguish it from biological neurons and point neurons. HTM is an acronym for Hierarchical Temporal Memory, a term used to describe our models of neocortex (Hawkins et al., 2011). HTM neurons used in the simulations for this paper have 128 dendrite/coincidence detectors with up to 40 synapses per dendrite. For clarity, Fig. 1C shows only a few dendrites and synapses.

## 2.2. Networks of Neurons Learn Sequences

Because all tissue in the neocortex consists of neurons with active dendrites and thousands of synapses, it suggests there are common network principles underlying everything the neocortex does. This leads to the question, what network property is so fundamental that it is a necessary component of sensory inference, prediction, language, and motor planning?

We propose that the most fundamental operation of all neocortical tissue is learning and recalling sequences of patterns (Hawkins and Blakeslee, 2004), what Karl Lashley famously called "the most important and also the most neglected problem of cerebral physiology" (Lashley, 1951). More specifically, we propose that each cellular layer in the neocortex implements a variation of a common sequence memory algorithm. We propose cellular layers use sequence memory for different purposes, which is why cellular layers vary in details such as size and connectivity. In this paper we illustrate what we believe is the basic sequence memory algorithm without elaborating on its variations.

We started our exploration of sequence memory by listing several properties required of our network in order to model the neocortex.

*1) On-line learning*
Learning must be continuous. If the statistics of the world change, the network should gradually and continually adapt with each new input.

*2) High-order predictions*
Making correct predictions with complex sequences requires the ability to incorporate contextual information from the past. The network needs to dynamically determine how much temporal context is needed to make the best predictions. The term "high-order" refers to "high-order Markov chains" which have this property.

*3) Multiple simultaneous predictions*
Natural data streams often have overlapping and branching sequences. The sequence memory therefore needs to make multiple predictions at the same time.

*4) Local learning rules*
The sequence memory must only use learning rules that are local to each neuron. The rules must be local in both space and time, without the need for a global objective function.

*5) Robustness*
The memory should exhibit robustness to high levels of noise, loss of neurons, and natural variation in the input. Degradation in performance under these conditions should be gradual.

All these properties must occur simultaneously in the context of continuously streaming data.

### 2.2.1. Mini-columns and Neurons: Two Representations

High-order sequence memory requires two simultaneous representations. One represents the feedforward input to the network and the other represents the feedforward input in a particular temporal context. To illustrate this requirement, consider two abstract sequences "ABCD" and "XBCY", where each letter represents a sparse pattern of activation in a population of neurons. Once these sequences are learned the network should predict "D" when presented with sequence "ABC" and it should predict "Y" when presented with sequence "XBC". Therefore, the internal representation during the subsequence "BC" must be different in the two cases; otherwise the correct prediction can't be made after "C" is presented.

Fig. 2 illustrates how we propose these two representations are manifest in a cellular layer of cortical neurons. The panels in Fig. 2 represent a slice through a single cellular layer in the neocortex (Fig. 2A). The panels are greatly simplified for clarity. Fig. 2B shows how the network represents two input sequences before the sequences are learned. Fig. 2C shows how the network represents the same input after the sequences are learned. Each feedforward input to the network is converted into a sparse set of active mini-columns. (Mini-columns in the neocortex span multiple cellular layers. Here we are only referring to the cells in a mini-column in one cellular layer.) All the neurons in a mini-column share the same feedforward receptive fields. If an unanticipated input arrives, then all the cells in the selected mini-columns will recognize the input pattern and become active. However, in the context of a previously learned sequence, one or more of the cells in the mini-columns will be depolarized. The depolarized cells will be the first to generate an action potential, inhibiting the other cells nearby. Thus a predicted input will lead to a very sparse pattern of cell activation that is unique to a particular element, at a particular location, in a particular sequence.

**[Figure 2 about here – see end of manuscript]**

### 2.2.2. Basal Synapses Are the Basis of Sequence Memory

In this theory, cells use their basal synapses to learn the transitions between input patterns. With each new feedforward input some cells become active via their proximal synapses. Other cells, using their basal synapses, learn to recognize this active pattern and upon seeing the pattern again, become depolarized, thereby predicting their own feedforward activation in the next input. Feedforward input activates cells, while basal input generates predictions. As long as the next input matches the current prediction, the sequence continues, Fig. 3. Fig. 3A shows both active cells and predicted cells while the network follows a previously learned sequence.

**[Figure 3 about here – see end of manuscript]**

Often the network will make multiple simultaneous predictions. For example, suppose that after learning the sequences "ABCD" and "XBCY" we expose the system to just the ambiguous sub-sequence "BC". In this case we want the system to simultaneously predict both "D" and "Y". Fig. 3B illustrates how the network makes multiple predictions when the input is ambiguous. The number of simultaneous predictions that can be made with low chance of error can again be calculated via Eq. (1). Because the predictions tend to be highly sparse, it is possible for a network to predict dozens of patterns simultaneously without confusion. If an input matches any of the predictions it will result in the correct highly-sparse representation. If an input does not match any of the predictions all the cells in a column will become active, indicating an unanticipated input.

Although every cell in a mini-column shares the same feedforward response, their basal synapses recognize different patterns. Therefore cells within a mini-column will respond uniquely in different learned temporal contexts, and overall levels of activity will be sparser when inputs are anticipated. Both of these attributes have been observed (Martin and Schröder, 2013; Vinje and Gallant, 2002; Yen et al., 2007).

For one of the cells in the last panel of Fig. 3A, we show three connections the cell used to make a prediction. In real neurons, and in our simulations, a cell would form 15 to 40 connections to a subset of a larger population of active cells.

### 2.2.3. Apical Synapses Create a Top-Down Expectation

Feedback axons between neocortical regions often form synapses (in layer 1) with apical dendrites of pyramidal neurons whose cell bodies are in layers 2, 3, and 5. It has long been speculated that these feedback connections implement some form of expectation or bias (Lamme et al., 1998). Our neuron model suggests a mechanism for top-down expectation in the neocortex. Fig. 4 shows how a stable feedback pattern to apical dendrites can predict multiple elements in a sequence all at the same time. When a new feedforward input arrives it will be interpreted as part of the predicted sequence. The feedback biases the input towards a particular interpretation. Again, because the patterns are sparse, many patterns can be simultaneously predicted.

**[Figure 4 about here – see end of manuscript]**

Thus there are two types of prediction occurring at the same time. Lateral connections to basal dendrites predict the next input, and top-down connections to apical dendrites predict multiple sequence elements simultaneously. The physiological interaction between apical and basal dendrites is an area of active research (Larkum, 2013) and will likely lead to a more nuanced interpretation of their roles in inference and prediction. However, we propose that the mechanisms shown in Figs. 2, 3 and 4 are likely to continue to play a role in that final interpretation.

### 2.2.4. Synaptic Learning Rule

Our neuron model requires two changes to the learning rules by which most neural models learn. First, learning occurs by growing and removing synapses from a pool of "potential" synapses (Chklovskii et al., 2004). Second, Hebbian learning and synaptic change occur at the level of the dendritic segment, not the entire neuron (Stuart and Häusser, 2001).

*Potential Synapses*

For a neuron to recognize a pattern of activity it requires a set of co-located synapses (typically fifteen to twenty) that connect to a subset of the cells that are active in the pattern to be recognized. Learning to recognize a new pattern is accomplished by the formation of a set of new synapses collocated on a dendritic segment.

Figure 5 shows how we model the formation of new synapses in a simulated HTM neuron. For each dendritic segment we maintain a set of "potential" synapses between the dendritic segment and other cells in the network that could potentially form a synapse with the segment (Chklovskii et al., 2004). The number of potential synapses is larger than the number of actual synapses. We assign each potential synapse a scalar value called "permanence" which represents stages of growth of the synapse. A permanence value close to zero represents an axon and dendrite with the potential to form a synapse but that have not commenced growing one. A 1.0 permanence value represents an axon and dendrite with a large fully formed synapse.

**[Figure 5 about here – see end of manuscript]**

The permanence value is incremented and decremented using a Hebbian-like rule. If the permanence value exceeds a threshold, such as 0.3, then the weight of the synapse is 1, if the permanence value is at or below the threshold then the weight of the synapse is 0. The threshold represents the establishment of a synapse, albeit one that could easily disappear. A synapse with a permanence value of 1.0 has the same effect as a synapse with a permanence value at threshold but is not as easily forgotten. Using a scalar permanence value enables on-line learning in the presence of noise. A previously unseen input pattern could be noise or it could be the start of a new trend that will repeat in the future. By growing new synapses, the network can start to learn a new pattern when it is first encountered, but only act differently after several presentations of the new pattern. Increasing permanence beyond the threshold means that patterns experienced more than others will take longer to forget.

HTM neurons and HTM networks rely on distributed patterns of cell activity, thus the activation strength of any one neuron or synapse is not very important. Therefore, in HTM simulations we model neuron activations and synapse weights with binary states. Additionally, it is well known that biological synapses are stochastic (Faisal et al., 2008), so a neocortical theory cannot require precision of synaptic efficacy. Although scalar states and weights might improve performance, they are not required from a theoretical point of view and all of our simulations have performed well without them. The formal learning rules used in our HTM network simulations are presented in the Materials and Methods section.

## 3. Simulation Results

Fig. 6 illustrates the performance of a network of HTM neurons implementing a high-order sequence memory. The network used in Fig. 6 consists of 2048 mini-columns with 32 neurons per mini-column. Each neuron has 128 basal dendritic segments, and each dendritic segment has up to 40 actual synapses. Because this simulation is designed to only illustrate properties of sequence memory it does not include apical synapses. The network exhibits all five of the desired properties for sequence memory listed earlier.

**[Figure 6 about here – see end of manuscript]**

Although we have applied HTM networks to many types of real-world data, in Fig. 6 we use an artificial data set to more clearly illustrate the network's properties. The input is a stream of elements, where every element is converted to a 2% sparse activation of mini-columns (40 active columns out of 2048 total). The network learns a predictive model of the data based on observed transitions in the input stream. In Fig. 6 the data stream fed to the network contains a mixture of random elements and repeated sequences. The embedded sequences are six elements long and require high-order temporal context for full disambiguation and best prediction accuracy, e.g. "XABCDE" and "YABCFG". For this simulation we designed the input data stream such that the maximum possible average prediction accuracy is 50% and this is only achievable by using high-order representations.

Fig. 6A illustrates on-line learning and high-order predictions. The prediction accuracy of the HTM network over time is shown in red. The prediction accuracy starts at zero and increases as the network discovers the repeated temporal patterns mixed within the random transitions. For comparison, the accuracy of a first-order network (created by using only one cell per column) is shown in blue. After sufficient learning, the high-order HTM network achieves the maximum possible prediction accuracy of 50% whereas the first-order network only achieves about 33% accuracy. After the networks reached their maximum performance the embedded sequences were modified. The accuracy drops at that point, but since the network is continually learning it recovers by learning the new high-order patterns.

Fig. 6B illustrates the robustness of the network. After the network reached stable performance we inactivated a random selection of neurons. At up to about 40% cell death there was minimal impact on performance. This robustness is due to the noise tolerance described earlier that occurs when a dendritic segment forms more synapses than necessary to generate an NMDA spike. At higher levels of cell death the network performance initially declines but then recovers as the network relearns the patterns using the remaining neurons.

## 4. Discussion

We presented a model cortical neuron that is substantially different than model neurons used in most artificial neural networks. The key feature of the model neuron is its use of active dendrites and thousands of synapses, allowing the neuron to recognize hundreds of unique patterns in large populations of cells. We showed that a neuron can reliably recognize many patterns, even in the presence of large amounts of noise and variation. In this model, proximal synapses define the feedforward receptive field of a cell. The basal and apical synapses depolarize the cell, representing predictions.

We showed that a network of these neurons will learn a predictive model of a stream of data. Basal synapses detect contextual patterns that predict the next feedforward input. Apical synapses detect feedback patterns that predict entire sequences. The operation of the neuron and the network rely on neural activity being sparse. The sequence memory model learns continuously, uses variable amounts of context to make predictions, makes multiple simultaneous predictions, relies on local learning rules, and is robust to failure of network elements, noise, and variation.

Although we refer to the network model as a "sequence memory", it is actually a memory of transitions. There is no representation or concept of the length of sequences or of the number of stored sequences. The network only learns transitions between inputs. Therefore, the capacity of a network is measured by how many transitions a given network can store. This can be calculated as the product of the expected duty cycle of an individual neuron (cells per column/column sparsity) times the number of patterns each neuron can recognize on its basal dendrites. For example, a network where 2% of the columns are active, each column has 32 cells, and each cell recognizes 200 patterns on its basal dendrites, can store approximately 320,000 transitions ($(32/0.02)*200$). The capacity scales linearly with the number of cells per column and the number of patterns recognized by the basal synapses of each neuron.

Another important capacity metric is how many times a particular input can appear in different temporal contexts without confusion. This is analogous to how many times a particular musical interval can appear in melodies without confusion, or how many times a particular word can be memorized in different sentences. If mini-columns have 32 cells it doesn't mean a particular pattern can have only 32 different representations. For example, if we assume 40 active columns per input, 32 cells per column, and one active cell per column, then there are $32^{40}$ possible representations of each input pattern, a practically unlimited number. Therefore, the practical limit is not representational but memory-based. The capacity is determined by how many transitions can be learned with a particular sparse set of columns.

So far we have only discussed cellular layers where all cells in the network can potentially connect to all other cells with equal likelihood. This works well for small networks but not for large networks. In the neocortex, it is well known that most regions have a topological organization. For example cells in region V1 receive feedforward input from only a small part of the retina and receive lateral input only from a local area of V1. HTM networks can be configured this way by arranging the columns in a 2D array and selecting the potential synapses for each dendrite using a 2D probability distribution centered on the neuron. Topologically organized networks can be arbitrarily large.

There are several testable predictions that follow from this theory.

1) The theory provides an algorithmic explanation for the experimentally observed phenomenon that overall cell activity becomes sparser during a continuous predictable sensory stream (Martin and Schröder, 2013; Vinje and Gallant, 2002; Yen et al., 2007). In addition, it predicts that unanticipated inputs will result in higher cell activity, which should be correlated vertically within mini-columns. Anticipated inputs on the other hand will result in activity that is uncorrelated within mini-columns. It is worth noting that mini-columns are not a strict requirement of this theory. The model only requires the presence of small groups of cells that share feedforward responses and that are mutually inhibitory. We refer to these groups as mini-columns, but the columnar aspect is not a requirement, and the groupings could be independent of actual mini-columns.

2) A second core prediction of the theory is that the current pattern of cell activity contains information about past stimuli. Early experimental results supporting this prediction have been reported in (Nikolić et al., 2009). Further studies are required to validate the exact nature of dynamic cell activity and the role of temporal context in high order sequences.

3) Synaptic plasticity should be localized to dendritic segments that have been depolarized via synaptic input followed a short time later by a back action potential. This effect has been reported (Losonczy et al., 2008), though the phenomenon has yet to be widely established.

4) There should be few, ideally only one, excitatory synapses formed between a given axon and a given dendritic segment. If an excitatory axon made many synapses in close proximity onto a single dendrite then the presynaptic cell would dominate in causing an NMDA spike. Two, three, or even four synapses from a single axon onto a single dendritic segment could be tolerated, but if axons routinely made more synapses to a single dendritic segment it would lead to errors. Pure Hebbian learning would seem to encourage forming multiple synapses. To prevent this from happening we predict the existence of a mechanism that actively discourages the formation of a multiple synapses after one has been established. An axon can form synapses onto different dendritic segments of the same neuron without causing problems, therefore we predict this mechanism will be spatially localized within dendritic segments or to a local area of an axonal arbor.

5) When a cell depolarized by an NMDA spike subsequently generates an action potential via proximal input, it needs to inhibit all other nearby excitatory cells. This requires a fast, probably single spike, inhibition. Fast-spiking basket inhibitory cells are the most likely source for this rapid inhibition (Hu et al., 2014).

6) All cells in a mini-column need to learn common feedforward responses. This requires a mechanism to encourage all the cells in a mini-column to become active simultaneously while learning feedforward patterns. This requirement for mutual excitation seems at odds with the prior requirement for mutual inhibition when one or more cells are slightly depolarized. We don't have a specific proposal for how these two requirements are met but we predict a mechanism where sometimes cells in a column are mutually excited and at other times they are mutually inhibited.

Pyramidal neurons are common in the hippocampus. Hence, parts of our neuron and network models might apply to the hippocampus. However, the hippocampus is known for fast learning, which is incompatible with growing new synapses, as synapse formation can take hours in an adult (Holtmaat and Svoboda, 2009; Knott et al., 2002; Niell et al., 2004; Trachtenberg et al., 2002). Rapid learning could be achieved in our model if instead of growing new synapses, a cell had a multitude of inactive, or "silent" synapses (Kerchner and Nicoll, 2008). Rapid learning would then occur by turning silent synapses into active synapses. The downside of this approach is a cell would need many more synapses, which is metabolically expensive. Pyramidal cells in hippocampal region CA2 have several times the number of synapses as pyramidal cells in neocortex (Megías et al., 2001). If most of these synapses were silent it would be evidence to suggest that region CA2 is also implementing a variant of our proposed sequence memory.

It is instructive to compare our proposed biological sequence memory mechanism to other sequence memory techniques used in the field of machine learning. The most common technique is Hidden Markov Models (HMMs) (Rabiner and Juang, 1986). HMMs are widely applied, particularly in speech recognition. The basic HMM is a first-order model and its accuracy would be similar to the first-order model shown in Fig. 6A. Variations of HMMs can model restricted high order sequences by encoding high-order states by hand. More recently, recurrent neural networks, specifically long short-term memory (LSTM) (Hochreiter and Schmidhuber, 1997), have become popular, often outperforming HMMs. Unlike HTM networks, neither HMMs nor LSTMs attempt to model biology in any detail; as such they provide no insights into neuronal or neocortical functions. The primary functional advantages of the HTM model over both these techniques are its ability to learn continuously, its superior robustness, and its ability to make multiple simultaneous predictions. A more detailed comparison can be found in S1 Table.

A number of papers have studied spiking neuron models (Ghosh-Dastidar and Adeli, 2009; Maass, 1997) in the context of sequences. These models are more biophysically detailed than the neuron models used in the machine learning literature. They show how spike-timing-dependent plasticity (STDP) can lead to a cell becoming responsive to a particular sequence of presynaptic spikes and to a specific time delay between each spike (Rao and Sejnowski, 2000; Ruf and Schmitt, 1997). These models are at a lower level of detail than the HTM model proposed in this paper. They explicitly model integration times of postsynaptic potentials and the corresponding time delays are typically sub-millisecond to a few milliseconds. They also typically deal with a very small subset of the synapses and do not explicitly model non-linear active dendrites. The focus of our work has been at a higher level. The work presented in this paper is a model of the full set of synapses and active dendrites on a neuron, of a networked layer of such neurons and the emergence of a computationally sophisticated sequence memory. An interesting direction for future research is to connect these two levels of modeling, i.e. to create biophysically detailed models that operate at the level of a complete layer of cells. Some progress is reported in (Billaudelle and Ahmad, 2015), but there remains much to do on this front.

A key consideration in learning algorithms is the issue of generalization, or the ability to robustly deal with novel patterns. The sequence memory mechanism we have outlined learns by forming synapses to small samples of active neurons in streams of sparse patterns. The properties of sparse representations naturally allow such a system to generalize. Two randomly selected sparse patterns will have very little overlap. Even a small overlap (such as 20%) is highly significant and implies that the representations share significant semantic meaning. Dendritic thresholds are lower than the actual number of synapses on each segment, thus segments will recognize novel but semantically related patterns as similar. The system will see similarity between different sequences and make novel predictions based on analogy.

Recently we showed that our sequence memory method can learn a predictive model of sensory-motor sequences (Cui et al., 2015). We also see it is likely that cortical motor sequences are generated using a variation of the same network model. Understanding how layers of cells can perform these different functions and how they work together is the focus of our current research.

## 5. Materials and Methods

Here we formally describe the activation and learning rules for an HTM sequence memory network. There are three basic aspects to the rules: initialization, computing cell states, and updating synapses on dendritic segments. These steps are described below, along with notation and some implementation details.

Notation: Let N represent the number of mini-columns in the layer, M the number of cells per column, and NM the total number of cells in the layer. Each cell can be in an active state, in a predictive (depolarized) state, or in a non-active state. Each cell maintains a set of segments each with a number of synapses. (In this figure we use the term "synapse" to refer to "potential synapses" as described in the body of the paper. Thus at any point in time some of the synapses will have a weight of 0 and some will have a weight of 1.) At any time step t, the set of active cells is represented by the M×N binary matrix $\mathbf{A}^t$, where $a_{ij}^t$ is the activity of the i'th cell in the j'th column. Similarly, the M×N binary matrix $\mathbf{\Pi}^t$ denotes cells in a predictive state at time t, where $\pi_{ij}^t$ is the predictive state of the i'th cell in the j'th column.

Each cell is associated with a set of distal segments, $\mathbf{D}_{ij}$, such that $\mathbf{D}_{ij}^d$ represents the d'th segment of the i'th cell in the j'th column. Each distal segment contains a number of synapses, representing lateral connections from a subset of the other NM − 1 cells. Each synapse has an associated permanence value (see Supplemental Fig. 2). Therefore, $\mathbf{D}_{ij}^d$ itself is also an M×N sparse matrix. If there are s potential synapses associated with the segment, the matrix contains s non-zero elements representing permanence values. A synapse is considered connected if its permanence value is above a connection threshold. We use $\widetilde{\mathbf{D}}_{ij}^d$ to denote a binary matrix containing only the connected synapses.

1) Initialization: the network is initialized such that each segment contains a set of potential synapses (i.e. with non-zero permanence value) to a randomly chosen subset of cells in the layer. The permanence values of these potential synapses are chosen randomly: initially some are connected (above threshold) and some are unconnected.

2) Computing cell states: All the cells in a mini-column share the same feed forward receptive fields. We assume that an inhibitory process has already selected a set of k columns that best match the current feed forward input pattern. We denote this set as $\mathbf{W}^t$. The active state for each cell is calculated as follows:

$$a_{ij}^t = \begin{cases} 1 & if j \in W^t \text{ and } \pi_{ij}^{t-1} = 1 \\ 1 & if j \in W^t \text{ and } \sum_i \pi_{ij}^{t-1} = 0 \\ 0 & otherwise \end{cases} \quad (2)$$

The first line will activate a cell in a winning column if it was previously in a predictive state. If none of the cells in a winning column were in a predictive state, the second line will activate all cells in that column. The predictive state for the current time step is then calculated as follows:

$$\pi_{ij}^t = \begin{cases} 1 & if \exists_d \left\| \widetilde{\boldsymbol{D}}_{ij}^d \circ \boldsymbol{A}^t \right\|_1 > \theta \\ 0 & otherwise \end{cases} \quad (3)$$

Threshold θ represents the NMDA spiking threshold and ∘ represents element-wise multiplication. At a given point in time, if there are more than θ connected synapses with active presynaptic cells, then that segment will be active (generate an NMDA spike). A cell will be depolarized if at least one segment is active.

3) *Updating segments and synapses*: the HTM synaptic plasticity rule is a Hebbian-like rule. If a cell was correctly predicted (i.e. it was previously in a depolarized state and subsequently became active via feedforward input), we reinforce the dendritic segment that was active and caused the depolarization. Specifically, we choose those segments $\boldsymbol{D}_{ij}^d$ such that:

$$\forall_{j \in W^t} \left( \pi_{ij}^{t-1} > 0 \right) \text{ and } \left\| \widetilde{\boldsymbol{D}}_{ij}^d \circ \boldsymbol{A}^{t-1} \right\|_1 > \theta \quad (4)$$

The first term selects winning columns that contained correct predictions. The second term selects those segments specifically responsible for the prediction.

If a winning column was unpredicted, we need to select one cell that will represent the context in the future if the current sequence transition repeats. To do this we select the cell with the segment that was closest to being active, i.e. the segment that had the most input even though it was below threshold. Let $\dot{\boldsymbol{D}}_{ij}^d$ denote a binary matrix containing only the positive entries in $\boldsymbol{D}_{ij}^d$. We reinforce a segment where the following is true:

$$\forall_{j \in W^t} \left( \sum_i \pi_{ij}^{t-1} = 0 \right) \text{ and} \quad (5)$$
$$\left\| \dot{\boldsymbol{D}}_{ij}^d \circ \boldsymbol{A}^{t-1} \right\|_1 = max_i \left( \left\| \dot{\boldsymbol{D}}_{ij}^d \circ \boldsymbol{A}^{t-1} \right\|_1 \right)$$

Reinforcing the above segments is straightforward: we wish to reward synapses with active presynaptic cells and punish synapses with inactive cells. To do that we decrease all the permanence values by a small value $p^-$ and increase the permanence values corresponding to active presynaptic cells by a larger value $p^+$:

$$\Delta \boldsymbol{D}_{ij}^d = p^+ \left( \dot{\boldsymbol{D}}_{ij}^d \circ \boldsymbol{A}^{t-1} \right) - p^- \dot{\boldsymbol{D}}_{ij}^d \quad (6)$$

The above rules deal with cells that are currently active. We also apply a very small decay to active segments of cells that did not become active. This can happen if segments were mistakenly reinforced by chance:

$$\Delta \boldsymbol{D}_{ij}^d = p^- \dot{\boldsymbol{D}}_{ij}^d \text{ where} \quad (7)$$
$$a_{ij}^t = 0 \text{ and } \left\| \widetilde{\boldsymbol{D}}_{ij}^d \circ \boldsymbol{A}^{t-1} \right\|_1 > \theta$$

The matrices $\Delta \boldsymbol{D}_{ij}^d$ are added to the current matrices of permanence values at every time step.

Implementation details: in our software implementation, we make some simplifying assumptions that greatly speed up simulation time for larger networks. Instead of explicitly initializing a complete set of synapses across every segment and every cell, we greedily create segments on a random cell and initialize potential synapses on that segment by sampling from currently active cells. This happens only when there is no match to any existing segment. In our simulations $N = 2048, M = 32, k = 40$. We typically connect between 20 and 40 synapses on a segment, and θ is around 15. Permanence values vary from 0 to 1 with a connection threshold of 0.5. $p^+$ and $p^-$ are small values that are tuned based on the individual dataset but typically less than 0.1. The full source code for the implementation is available on Github at https://github.com/numenta/nupic

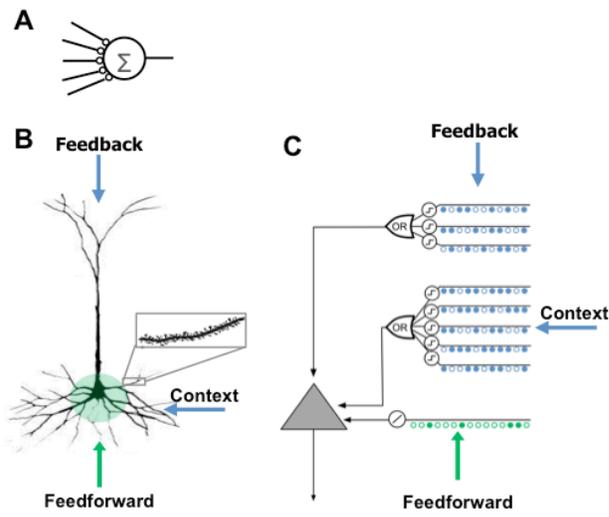

**Figure 1:** Comparison of neuron models. **A**) The neuron model used in most artificial neural networks has few synapses and no dendrites. **B**) A neocortical pyramidal neuron has thousands of excitatory synapses located on dendrites (inset). The co-activation of a set of synapses on a dendritic segment will cause an NMDA spike and depolarization at the soma. There are three sources of input to the cell. The feedforward inputs (shown in green) which form synapses proximal to the soma, directly lead to action potentials. NMDA spikes generated in the more distal basal and apical dendrites depolarize the soma but typically not sufficiently to generate a somatic action potential. **C**) An HTM model neuron models dendrites and NMDA spikes with an array of coincident detectors each with a set of synapses (only a few of each are shown).

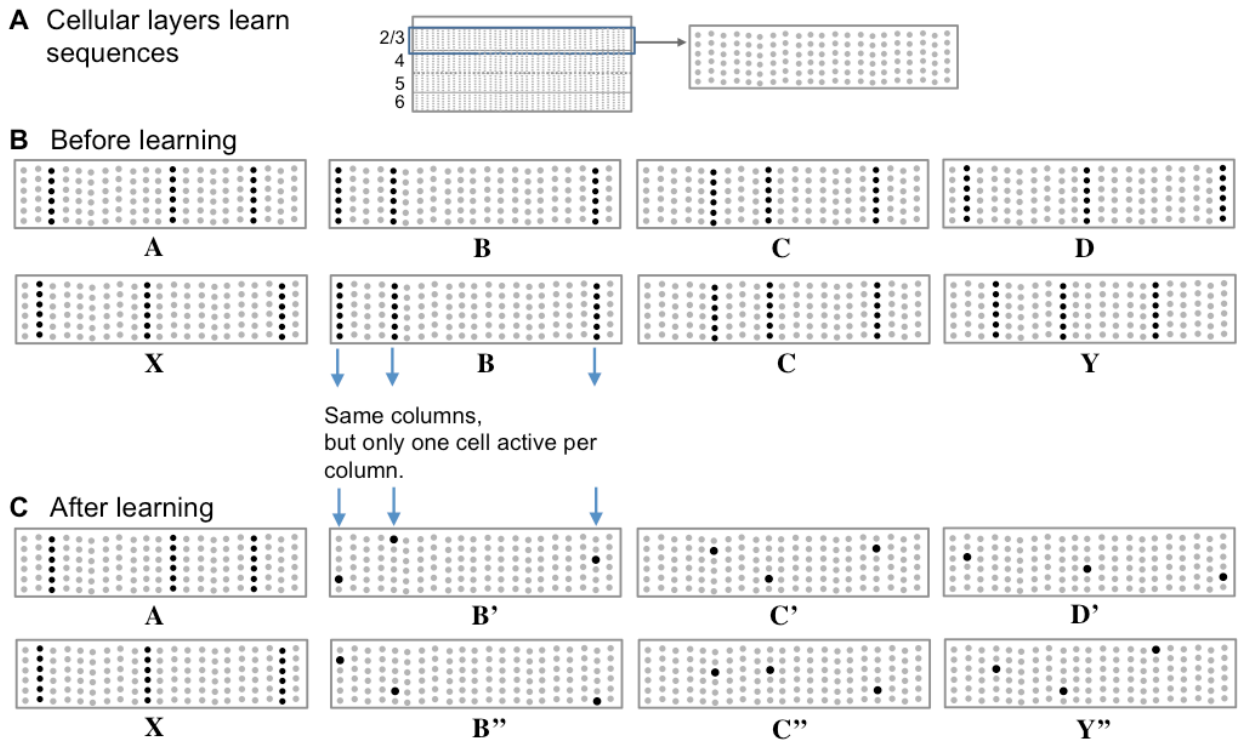

**Figure 2**: Representing sequences in cortical cellular layers. **A)** The neocortex is divided into cellular layers. The panels in this figure show part of one generic cellular layer. For clarity, the panels only show 21 mini-columns with 6 cells per column. **B)** Input sequences ABCD and XBCY are not yet learned. Each sequence element invokes a sparse set of mini-columns, only three in this illustration. All the cells in a mini-column become active if the input is unexpected, which is the case prior to learning the sequences. **C)** After learning the two sequences, the inputs invoke the same mini-columns but only one cell is active in each column, labeled B', B'', C', C'', D' and Y''. Because C' and C'' are unique, they can invoke the correct high-order prediction of either Y or D.

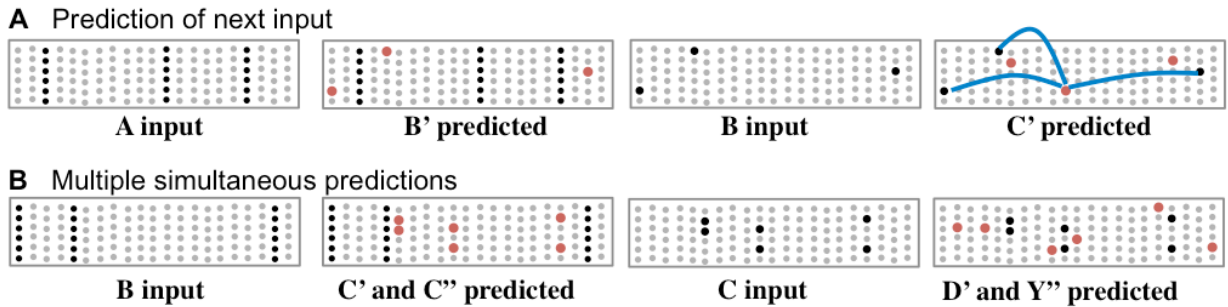

**Figure 3:** Basal connections to nearby neurons predict the next input. **A)** Using one of the sequences from Fig. 2, both active cells (black) and depolarized/predicted cells (red) are shown. The first panel shows the unexpected input A, which leads to a prediction of the next input B' (second panel). If the subsequent input matches the prediction then only the depolarized cells will become active (third panel), which leads to a new prediction (fourth panel). The lateral synaptic connections used by one of the predicted cells are shown in the rightmost panel. In a realistic network every predicted cell would have 15 or more connections to a subset of a large population of active cells. **B)** Ambiguous sub-sequence "BC" (which is part of both ABCD and XBCY) is presented to the network. The first panel shows the unexpected input B, which leads to a prediction of both C' and C''. The third panel shows the system after input C. Both sets of predicted cells become active, which leads to predicting both D and Y (fourth panel). In complex data streams there are typically many simultaneous predictions.

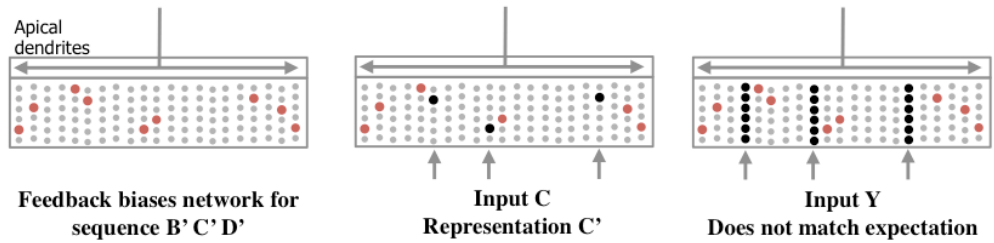

**Figure 4:** Feedback to apical dendrites predicts entire sequences.

This figure uses the same network and representations as Fig. 2. Area labeled "apical dendrites" is equivalent to layer 1 in neocortex; the apical dendrites (not shown) from all the cells terminate here. In the figure, the following assumptions have been made. The network has previously learned the sequence ABCD as was illustrated in Fig. 2. A constant feedback pattern was presented to the apical dendrites during the learned sequence, and the cells that participate in the sequence B'C'D' have formed synapses on their apical dendrites to recognize the constant feedback pattern.

After the feedback connections have been learned, presentation of the feedback pattern to the apical dendrites is simultaneously recognized by all the cells that would be active sequentially in the sequence. These cells, shown in red, become depolarized (left pane). When a new feedforward input arrives it will lead to the sparse representation relevant to the predicted sequence (middle panel). If a feedforward pattern cannot be interpreted as part of the expected sequence (right panel) then all cells in the selected columns become active indicative of an anomaly. In this manner apical feedback biases the network to interpret any input as part of an expected sequence and detects if an input does not match any one of the elements in the expected sequence.

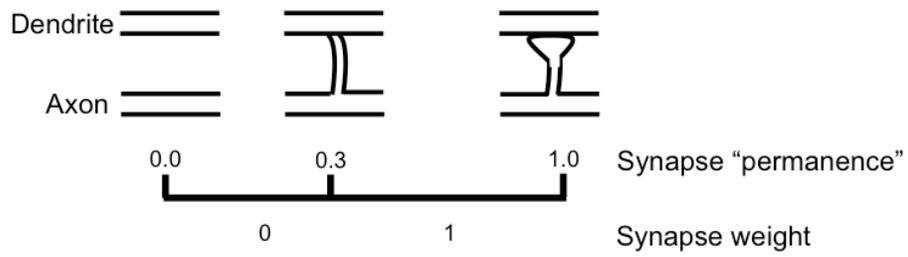

**Figure 5:** Learning by growing new synapses. Learning in an HTM neuron is modeled by the growth of new synapses from a set of potential synapses. A "permanence" value is assigned to each potential synapse and represents the growth of the synapse. Learning occurs by incrementing or decrementing permanence values. The synapse weight is a binary value set to 1 if the permanence is above a threshold.

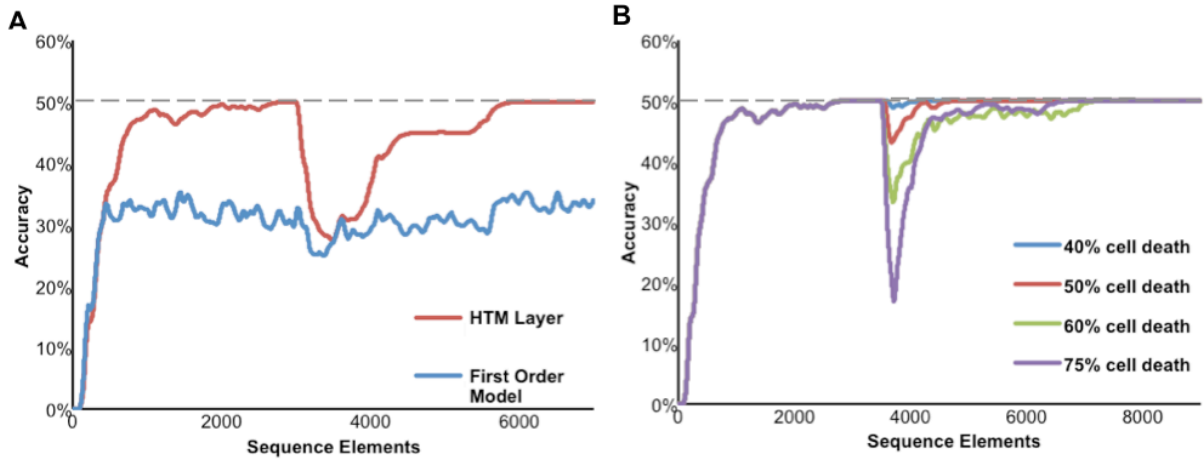

**Figure 6:** Simulation results of the sequence memory network. The input stream used for this figure contained high-order sequences mixed with random elements. The maximum possible average prediction accuracy of this data stream is 50%. **A)** High-order on-line learning. The red line shows the network learning and achieving maximum possible performance after about 2500 sequence elements. At element 3000 the sequences in the data stream were changed. Prediction accuracy drops and then recovers as the model learns the new temporal structure. For comparison, the lower performance of a first-order network is shown in blue. **B)** Robustness of the network to damage. After the network reached stable performance we inactivated a random selection of neurons. At up to 40% cell death there is almost no impact on performance. At greater than 40% cell death the performance of the network declines but then recovers as the network relearns using remaining neurons.

**S1 Text. Chance of Error When Recognizing Large Patterns with a Few Synapses**

**Formula for calculating chance of error**

A non-linear dendritic segment can robustly classify a pattern by sub-sampling (forming synapses to) a small number of cells from a large population. Assuming a random distribution of patterns, the exact probability of a false match, following is given by the following equation:

$$\frac{\sum_{b=\theta}^{s} \binom{s}{b} \times \binom{n-s}{a-b}}{\binom{n}{a}}$$

$n$ = cell population size
$a$ = number of active cells
$s$ = number of synapses on segment
$\theta$ = NMDA spike threshold

**Table A: Chance of error due to sub-sampling**

This table demonstrates the effect of sub-sampling on the probability of a false match using the above equation. The chance of an error drops rapidly as the sampling size increases. A small number of synapses is sufficient for reliable matching.

| $s$ | Probability of false match |
|---|---|
| 6 | $9.9 \times 10^{-13}$ |
| 8 | $9.8 \times 10^{-17}$ |
| 10 | $9.8 \times 10^{-21}$ |

$n = 200{,}000$
$a = 2{,}000$
$\theta = s$

**Table B: Chance of error with addition of 50% noise immunity**

This table demonstrates robustness to noise. By forming more synapses than required for an NMDA spike, a neuron can be robust to large amounts of noise and pattern variation and still have low probability of a false match. For example, with $s = 2\theta$ the system will be immune to 50% noise. The chance of an error drops rapidly as $\theta$ increases; even with noise a small number of synapses is sufficient for reliable matching.

| $\theta$ | $s$ | Probability of false match |
|---|---|---|
| 6 | 12 | $8.7 \times 10^{-10}$ |
| 8 | 16 | $1.2 \times 10^{-12}$ |
| 10 | 20 | $1.6 \times 10^{-15}$ |
| 12 | 24 | $2.3 \times 10^{-18}$ |

$n = 200{,}000$
$a = 2{,}000$

**Table C: Chance of error with addition of mixing synapses on a dendritic segment**

This table demonstrates that mixing synapses for $m$ different patterns on a single dendritic segment will still not cause unacceptable errors. By setting $s = 2m\theta$ we can see how a segment can recognize $m$ independent patterns and still be robust to 50% noise. It is possible to get very high accuracy with larger $m$ by using a slightly higher threshold.

| $\theta$ | $m$ | $s$ | Probability of false match |
|---|---|---|---|
| 10 | 2 | 40 | $6.3 \times 10^{-12}$ |
| 10 | 4 | 80 | $8.5 \times 10^{-9}$ |
| 10 | 6 | 120 | $4.2 \times 10^{-7}$ |
| 15 | 6 | 120 | $1.7 \times 10^{-12}$ |

$n = 200{,}000$
$a = 2{,}000$

|  | **HTM** | **HMMs** | **LSTM** |
|---|---|---|---|
| High order sequences | Yes | Limited | Yes |
| Discovers high order sequence structure | Yes | No | Yes |
| Local learning rules | Yes | No | No[*] |
| Continuous learning | Yes | No | No |
| Multiple simultaneous predictions | Yes | No | No |
| Unsupervised learning | Yes | Yes | No |
| Robustness and fault tolerance | Very high | No | Yes |
| Detailed mapping to neuroscience | Yes | No | No |
| Probabilistic model | No | Yes | No |

**S1 Table, Comparison of Common Sequence Memory Algorithms**

Table comparing two common sequence memory algorithms (HMM and LSTM) to proposed model (HTM).
 * Although weight updated rules are local, LSTMs require computing a global error signal that is then back propagated.